\documentclass{aastex631}

\usepackage{booktabs}
\usepackage{soul}

\newcommand{\ang}[1]{\ensuremath{{#1}^{\circ}}}
\newcommand{\is}{\textit{in situ}}

\newcommand{\ie}{\textit{i.e.}}
\newcommand{\eg}{\textit{e.g.}}

\shortauthors{Regnault et al.}
\graphicspath{{./}{figures/}}
\begin{document}

\title{Investigating the Magnetic Structure of Interplanetary Coronal Mass Ejections using Simultaneous Multi-Spacecraft In situ Measurements}

%Quantifying and Reducing the Observed Ageing Effect of CMEs through Simultaneous In-situ Measurements}
\author[0000-0002-4017-8415]{F. Regnault}
\author[0000-0002-0973-2027]{N. Al-Haddad}
\author[0000-0002-1890-6156]{N. Lugaz}
\author[0000-0001-8780-0673]{C.~J. Farrugia}
\author[0000-0002-2917-5993]{W. Yu}
\affiliation{Space Science Center, Institute for the Study of Earth, Oceans, and Space, University of New Hampshire}
\author[0000-0001-9992-8471]{E.~E.~Davies}
\affiliation{Space Science Center, Institute for the Study of Earth, Oceans, and Space, University of New Hampshire}
\affiliation{Austrian Space Weather Office, GeoSphere Austria, Graz, Austria}
\author[0000-0003-3752-5700]{A.~B. Galvin}
\affiliation{Space Science Center, Institute for the Study of Earth, Oceans, and Space, University of New Hampshire}
\author[0000-0002-5996-0693]{B.~Zhuang}
\affiliation{Space Science Center, Institute for the Study of Earth, Oceans, and Space, University of New Hampshire}

\begin{abstract}
\textit{In situ} measurements from spacecraft typically provide a time series
at a single location through coronal mass ejections (CMEs) and they have been one of the main methods to investigate CMEs.
CME properties derived from these \is\ measurements are affected by temporal changes that occur as the CME passes over the spacecraft, such as radial expansion and ageing, as well as spatial variations within a CME.
This study uses multi-spacecraft measurements of the same CME at close separations to investigate both the spatial variability (how different a CME profile is when probed by two spacecraft close to each other) and the so-called ageing effect (the effect of the time evolution on \is\ properties).
We compile a database of 19 events from the past four decades measured by  two spacecraft with a radial separation $<0.2$ au and an angular separation $<\ang{10}$. We find that the average magnetic field strength measured by the two spacecraft differs by 18\% of the typical average value, which highlights non-negligible spatial or temporal variations. 
For one particular event, measurements taken by the two spacecraft allow us to quantify and significantly reduce the ageing effect to estimate the asymmetry of the magnetic field strength profile. This study reveals that single-spacecraft time series near 1~au can be strongly affected by ageing and that correcting for self-similar expansion does not capture the whole ageing effect.
%CME properties derived from such measurements, even corrected for self-similar expansion, may not always correspond to the ``true'' CME time-independent properties.
%This \mod{pilot} study also highlights the need for multi-spacecraft missions with small radial separations dedicated to the study of CMEs to address the effects of ageing and to investigate the structure of the magnetic ejecta.

\end{abstract}

%As they propagate, these magnetic ejectas interact with the coronal and interplanetary magnetic field and plasma, and are referred to as coronal mass ejections (CMEs).


\section{Introduction}

%Presentation of what is an ICME. Complex magnetic structures.
Magnetic instabilities occurring in the solar atmosphere can lead to the ejection of magnetic
structures %(called magnetic ejecta, ME) 
that subsequently propagate through the corona and the heliosphere.
Such events are referred to as coronal mass ejections (CMEs). CMEs have been imaged in the corona by white-light coronagraphs since the early 1970s \citep[]{tousey1973} and are now routinely imaged by SOHO/LASCO \citep[]{brueckner1995} and STEREO/SECCHI/COR1,2 \citep[]{howard2008} coronagraphs. In addition,
\is\ instruments on board interplanetary probes directly measure  
the magnetic and plasma properties of CMEs as they pass over the spacecraft. 
When measured \is, CMEs are typically composed of a dense and hot sheath, preceded about half of the time by a fast-forward shock \citep{jian2006,salman2020b}, and a low-proton $\beta$, high magnetic field, low temperature magnetic ejecta (ME) \citep{regnault2020,salman2020b}. Based on such measurements of the same event by five spacecraft, \cite{burlaga1981} identified magnetic clouds (MCs) as one type of magnetic ejecta that have clearly defined properties, \ie,  a rotation of the magnetic field vector, strong magnetic field strength, and a low proton temperature. However, more than half the events measured \is\ at 1~au do not have these signatures \citep[]{gosling1990,richardson2010}. In this study we use the  term ``magnetic ejecta'' to denote the ejecta-part of the CME, irrespective of whether or not it is a magnetic cloud. 
Traditionally, there has been a distinction between CMEs observed remotely in the corona and interplanetary CMEs, or ICMEs, measured in situ by spacecraft. However, since the 2000s, it has been possible to remotely observe “ICMEs” in the interplanetary space, first with SMEI \citep{eyles2003}, then STEREO/HI \citep{harrison2009}, and more recently to measure CMEs in situ in the upper corona and innermost heliosphere with Parker Solar Probe and Solar Orbiter. Both of these missions combine in situ measurements and remote observations, often simultaneously, of CMEs/ICMEs. As such, the distinction between CMEs and ICMEs has become smeared and harder to justify. Throughout this paper, we use the term “CME” to describe the whole ejection, irrespective of how and where it is measured. In doing so, we follow the conventions of past researchers during the last decade (\eg, see \citealt{howard2012,lugaz2015}). This also follows decades of research which have clearly identified the CMEs measured in the corona as the source of the CMEs measured in situ (\eg, see \citealt{bothmer1998}).

Most of the time, \is\ measurements of CMEs are from a single spacecraft. This provides the equivalent to a one-dimensional cut through the 3D CME structure under the assumption of a CME being a static structure. The limitations of such measurements make it  difficult to investigate the global properties of the 3D structure of CMEs. 
For this reason, fitting and reconstruction techniques, like the Grad-Shafranov technique \citep{hu2002} or the Lundquist analytical model \citep{lundquist1951,burlaga1988, lepping1990}, have been developed to deduce the global properties of the CME using a single-spacecraft \is\ time-series. Such techniques, however,  require strong assumptions, such as the presence of an invariance direction and/or assumptions on the shape of the magnetic structure, that are not always justified, and that typically cannot be tested. Most often, the ME structure is assumed to be that of a highly twisted magnetic flux rope and even other types of structures may be reconstructed as such \citep{al-haddad2011,al-haddad2019a}.  In addition, different techniques are often found to be inconsistent with one another \citep{riley2004a,al-haddad2013,martinic2022}.
For this reason, in this study, we limit the number of assumptions that we make, staying away from fitting techniques, and try to interpret the data without relying on models as much as possible.
Our main assumption in this study is that the ME is a coherent magnetic structure up to an angular separation of \ang{15-20} (\citealt{owens2017,lugaz2018}). Moreover, we emphasize that we use the more general definition of ME throughout this study, and therefore, do not assume that the ME must have a magnetic flux rope structure or correspond to a magnetic cloud being crossed away from its center. Thus, we do not assume that there is an invariant direction in the MEs we analyze.

Sometimes, the same CME is observed at different locations by two (or rarely, three) spacecraft. We refer to this instance as a measurement with spacecraft in conjunction. In the case of an almost radial alignment, the evolution of the CME properties can be investigated \citep{leitner2007,nakwacki2011,good2019,salman2020a,davies2022}. Such conjunctions are relatively rare in the inner heliosphere ($\sim$ 0.3 to 1.5~au). Most studies are limited by the required assumption that the same part of the CME is being observed. In addition, the evolution of the CME due to its interactions with the solar wind and corotating structures \citep{winslow2016,davies2020} cannot be easily distinguished from the internal evolution of the CME \citep{lugaz2020a,davies2021a}. 

An additional difficulty is that the CME continues to evolve as it passes over a spacecraft. Near 1~au, a ME duration is typically around 20 hours \citep{nieves-chinchilla2018}, which is about 25\% of its age.
 The effect of the temporal evolution of the CME properties over the duration of the spacecraft crossing is referred to as the ageing effect \citep{osherovich1993}. It causes difficulties to compute and derive some of the properties of the ME \is\ profile, such as its average magnetic field strength, the asymmetry of its profile, among others. As noted previously by \cite{klein1982}, the speed profile of MEs as measured \is\ is often decreasing as it passes over the spacecraft, which is typically interpreted as a result of ME expansion and can be corrected using assumptions regarding the expansion \citep{farrugia1993}. \cite{demoulin2020} developed a methodology to deduce the instantaneous magnetic field strength  profile of MEs by quantifying the expansion using a fit of the speed within the ME assuming a self-similar expansion. Their methodology assumes that all the change in the velocity as the ME passes over the spacecraft is due to expansion, and correct for it. This method neglects the changes due to the global evolution of the ME, e.g., the ME center speed may decrease during the time taken for the ME to pass over the spacecraft. Hereafter, we distinguish between changes due to expansion and those due to ageing, as described in \cite{osherovich1993}.

%\cf{Reference: Osherovich, V.A., C. J. Farrugia, and L.F. Burlaga, Dynamics of aging
%magnetic clouds,  Adv. Space Res., 13, 6(6), 57, 1993.} 
%
%\cf{NOTE: IN this paper we say: "Therefore the asymmetry of the total field strength profile is a consequence of aging and expansion of the magnetic cloud"...see their Figure 3}
%

Rarely, the same CME is probed simultaneously by more than one spacecraft \citep{farrugia2011a,kilpua2011,winslow2021,lugaz2022}. When this is the case, the CME is typically measured by two spacecraft at approximately the same distance but separated by an azimuthal angle ranging from a fraction of a degree to 40--50$^\circ$. With such spacecraft configurations, it is possible to gain direct information about the global structure of MEs, for example whether the assumption of axial invariance is valid \citep{mostl2009b,mostl2009a,mulligan2013}, but such conjunctions have been limited to few cases \citep{lugaz2018}.

There has not been, to the best of our knowledge, any study looking at simultaneous measurements of CMEs by two (or more) spacecraft at different radial distances but approximately aligned in longitude. This configuration is the same as for investigations of CME evolution discussed above but for spacecraft in relatively close radial proximity. Since a ME is typically about 0.2~au in radial extent at 1~au \citep{lepping1990}, we select events where spacecraft are radially separated by less than 0.2~au.  Such measurements enable, in the case of a radial alignment, the investigation of the CME temporal evolution (ageing) rather than its radial evolution (propagation).
These types of simultaneous measurements on which we focus here can significantly improve our understanding of the complex magnetic structure of MEs and can help us, for instance, to quantify the ageing and expansion processes occurring during the ME crossing. We discuss further the insights gained by the different spacecraft configurations in Section \ref{sec:discussion}.

The aim of this paper is to take advantage of such simultaneous measurements. We do so, first, by investigating how consistent ME properties are when measured simultaneously by two spacecraft with small radial separation and, second, by quantifying the ageing effect and proposing ways to reduce it. The latter allows us to deduce instantaneous CME properties and thus have a better understanding of the ME structure. 

To find such events measured simultaneously by two spacecraft with small but non-negligible radial separations and small angular separations (that includes latitudinal and longitudinal separations), we search more than four decades of data for instances where two spacecraft measured the same CME at the same time.  In Section~\ref{sec:catalog}, we describe our procedure to identify such events. These events are referred to as ``simultaneous measurements''. In Section~\ref{sec:disc_ME} we first show an example of a simultaneous measurement event and then analyze all the events at our disposal in terms of typical differences between the two spacecraft. We focus on the magnetic field strength and the asymmetry of its temporal profile. This enables us to conclude that the differences are not primarily due to radial propagation, but must be a combination of temporal changes (ageing) and geometrical effects. In Section~\ref{sec:ageing}, we present a case study of a simultaneous measurement, allowing us to quantify for the first time the effect of ageing on \is\ profiles. We also use this case to investigate two methods to reduce this effect.
In Section~\ref{sec:discussion}, we show what can be done given different spacecraft configuration during simultaneous measurements.Finally, in Section~\ref{sec:conclusion} we conclude and discuss our findings.

%\mod{
%Traditionally, there had been a distinction between CMEs observed remotely in the corona and interplanetary CMEs, or ICMEs, measured \is\ by spacecraft. However, since the 2000s, it has been possible to remotely observe ``ICMEs'' in the interplanetary space, first with SMEI \citep[]{eyles2003}, then STEREO/HI \citep[]{harrison2009}, and more recently to measure CMEs \is\ in the upper corona and innermost heliosphere with Parker Solar Probe and Solar Orbiter. Both of these missions combine \is\ measurements and remote observations, often simultaneously, of CMEs/ICMEs. As such, the distinction between CMEs and ICMEs has become harder to justify. Throughout this paper, we use the term ``CME'' to describe the whole ejection, irrespective of how and where it is measured. In doing so, we follow the conventions of past researchers during the last decade \citep[e.g., see][]{howard2012,lugaz2015}. This also follows decades of research which have clearly identified the CMEs measured in the corona as the source of the CMEs measured \is\ \citep[e.g., see][]{bothmer1998}.
%}

\section{Simultaneous Measurement Catalog} %Close conjunctions catalog}
\label{sec:catalog}

We use existing catalogs of CMEs measured in the inner heliosphere  to identify events measured by two spacecraft with moderate separations, and use an automated algorithm --see below for quantitative explanations-- to search the catalog for such events. 
Combined, these catalogs represent a total of 2374 potential CMEs spread over more than four decades of data acquisition by eleven different interplanetary probes. Some CMEs may appear in multiple lists; for example in the list of CMEs measured by ACE of \cite{richardson2010} and in the HELIO4CAST ICMECAT list of \cite{mostl2020} covering CMEs measured by Wind. However, if a duplicate of the same CME is found after all of the filtering processes described in the next section, we manually remove it from the list of potential simultaneous measurements.

The list of CME catalogs used in this study is the following:

%\begin{table}[h]
%\centering
%\begin{tabular}{c|c|c}
%    List & Spacecraft & \# ICMEs \\
%     \hline
%     \cite{mostl2017,mostl2020} & Bepi, MESSENGER, L1, SA,SB & 1037 \\
%     \cite{richardson2010}& L1 & 498 \\
%     %\cite{lee2017} & L1 & 10 \\
%     \cite{wang2005} & H1, H2, PVO & 596\\
%     \cite{cane1997} & H1, H2 & 29\\
%     \cite{davies2021} & L1, Juno & 27 \\
%     \cite{mulligan1999} & NEAR, Wind & 1 \\
%     \cite{jian2008} & PVO & 124
%\end{tabular}
%\label{tab:ICME_list}
%\caption{\fr{can be turned into a list if needed.}}
%\end{table}

\begin{itemize}
    \item the combined HELIO4CAST ICMECAT\footnote{\url{https://helioforecast.space/icmecat}}\footnote{\url{https://doi.org/10.6084/m9.figshare.6356420.v12}} catalog of CMEs of \cite{mostl2017,mostl2020} where we focused on events measured by BepiColombo, Parker Solar Probe, Solar Orbiter, MESSENGER, ACE/Wind, STEREO-A, and STEREO-B / 1073 CMEs / from 0.05 to 1 au / 1995 to 2021
    \item \cite{richardson2010} ACE / 531 CMEs / at 1 au / 1996 to 2022
     %\cite{lee2017} , L1 , 10 \\
\item \cite{wang2005}, Helios-1, Helios-2, Pioneer Venus Orbiter (PVO), ACE/ 596 CMEs /  from 0.3 to 1 au / 1975 to 2003
    \item \cite{cane1997}, Helios-1, Helios-2 / 29 CMEs / from 0.3 to 1 au / 1975 to 1978
    \item \cite{davies2022}, Wind, Juno / 27 conjunctions / from 1 to 2 au / 2011 to 2014
    \item \cite{mulligan1999}, NEAR (Near Earth Asteroid Rendezvous), Wind / 4 conjunctions / from 1 to 1.6 au / 1997
    \item \cite{jian2008}, PVO / 124 CMEs / from 0.72 to 0.73 au au / 1979 to 1988
\end{itemize}

%\subsection{Finding the close conjunctions}

To identify CMEs measured by two spacecraft with close spatial separations, we use the position (using the interpolated daily position obtained at \url{https://omniweb.gsfc.nasa.gov/coho/helios/heli.html}) of the spacecraft at the time of a detected CME and we look for the position of the other spacecraft. We take the trajectory of the Earth for all spacecraft at L1 and we take the Venus trajectory for the PVO mission once it entered orbit around Venus in December 1978 \citep{colin1980}.

If, during a CME encounter with a spacecraft, another spacecraft has a radial separation of less than 0.2~au and an angular separation ($\Delta \alpha$) lower than \ang{10}, we then flag this event as a potential simultaneous measurement. 
The angular separation is the angle between the two Sun-spacecraft lines in 3D space. It is dominated by the longitudinal separation between the spacecraft since most spacecraft orbit in the ecliptic plane, but includes also the latitudinal separation. We compute it using the dot product of the two spacecraft position vectors.
We do not set a lower limit for the radial separation because it is difficult to determine a threshold value related to the CME radial size at different distances from the Sun and we do not want to exclude possible interesting cases. We instead remove manually the events for which the two spacecraft are too close (see details below). 
For CMEs detected at L1 after 1995, we use ACE or Wind data, when they are available and depending on which spacecraft the CME has been reported on in the database (all CMEs should be observed by both, except when Wind was inside Earth's magnetosphere during the first few years of the mission). For CMEs before 1995, we use the IMP-8 spacecraft. We do not consider cases where the CME was only measured by spacecraft at or near L1, for example, simultaneous measurements by ACE, Wind and IMP-8, since the separation between the spacecraft is too small to have any effects on the interpretation of the large-scale structure of MEs. For example, \citet{lugaz2018} showed that the measurements are nearly identical up to separations of $\sim 200~R_E$ (i.e. $\sim 0.4^\circ$).

We find 126 potential simultaneous measurements satisfying these criteria for the radial and angular separations. This initial list does not take into consideration whether data is available at both spacecraft. They are found using only the positions of the spacecraft. A significant portion of these potential simultaneous measurements occur during the early cruise phase of planetary missions, when data may not be  available. %In particular, we don't have any data for the early cruise phase of BepiColombo or MESSENGER.

In a second step, we visually inspect these 126 events and remove all those for which there is no available data at one of the spacecraft or cases where the two spacecraft are too close (for example, the 2006 December 13 CME measured by the two STEREO spacecraft as well as ACE and Wind just after the launch of the STEREO spacecraft). We then confirm that there is temporal overlap between the measurements since for example, a small CME may not be measured simultaneously by two spacecraft separated by 0.2~au. We also removed potential events where the CME signatures (mainly low proton $\beta$ and a smooth magnetic field) are clear at only one of the spacecraft. 

Table \ref{tab:boundaries} presents the catalog of the final 19 CME events for which there are multi-spacecraft simultaneous measurements. 
We generally use the ME boundaries as provided in existing catalogs except for some MEs for which we adjusted the boundaries following visual inspection so that MEs at both spacecraft have consistent boundaries.
Thus, for the measurements at the second spacecraft, we attempt to match the features observed at the first spacecraft profile with those of the second one. Note that, in almost all cases, we only have plasma data at one of the spacecraft, making it important to define boundaries primarily through discontinuities in the magnetic field vector. We would like to note that the MEs presented in this catalog are complicated, and with the lack of plasma measurements in most of the MEs, the boundaries can be difficult to identify clearly. %Therefore, timestamps listed in Table \ref{tab:boundaries} should be treated with caution.}

\begin{table}[h]
\centering
\footnotesize
\begin{tabular}{cllcccccccc}
\toprule
 \# &        SC1 &      SC2 &   r1 &   r2 &        ME start1 &          ME end1 &        ME start2 &          ME end2 & overlap & $\Delta \alpha$ \\

        &  & & [au] & [au] & & & & & [\%] &  \ang{} \\
\midrule
1 & SolarOrbiter &     Wind & 0.84 & 0.99 & 2021-11-04T07:09 & 2021-11-04T19:38 & 2021-11-04T10:52 & 2021-11-05T04:14 &    70.3 & 2.2 \\
2 & Helios-2 &      IMP & 0.94 & 0.98 & 1978-01-04T07:07 & 1978-01-05T14:51 & 1978-01-04T10:54 & 1978-01-05T16:00 &    88.1 & 5.6 \\
3 &      IMP & Helios-2 & 0.98 & 0.95 & 1978-01-06T01:06 & 1978-01-06T16:53 & 1978-01-06T01:16 & 1978-01-06T14:47 &    99.0 & 6.2 \\
4 & Helios-1 &      IMP & 0.86 & 0.99 & 1975-11-17T04:45 & 1975-11-17T18:08 & 1975-11-17T12:02 & 1975-11-17T23:35 &    45.5 & 5.1 \\
5 & Wind &     NEAR & 1.00 & 1.18 & 1997-12-10T18:35 & 1997-12-12T00:04 & 1997-12-11T18:46 & 1997-12-12T23:33 &    18.0 & 1.2 \\
6 &    Wind &     Juno & 1.00 & 1.07 & 2011-09-17T15:35 & 2011-09-18T08:51 & 2011-09-17T20:00 & 2011-09-18T14:43 &    74.4 & 6.6 \\
7 &    Wind &     Juno & 0.99 & 1.11 & 2011-09-26T20:54 & 2011-09-28T14:14 & 2011-09-27T09:54 & 2011-09-28T23:30 &    68.5 & 7.7 \\
8 & Wind &     Juno & 0.99 & 1.14 & 2011-10-05T09:56 & 2011-10-07T02:06 & 2011-10-05T21:32 & 2011-10-07T10:22 &    71.1 & 6.8 \\
9 &    Wind &     Juno & 0.98 & 1.10 & 2013-10-30T17:34 & 2013-10-31T05:00 & 2013-10-31T02:32 & 2013-10-31T23:47 &    21.6 & 4.9 \\
10 &      Wind &     Juno & 0.98 & 1.18 & 2013-11-11T17:44 & 2013-11-12T01:56 & 2013-11-11T22:58 & 2013-11-12T08:06 &    36.1 & 4.5 \\
11 & STEREO-A &      PSP & 0.96 & 0.92 & 2019-11-02T21:19 & 2019-11-03T22:02 & 2019-11-02T21:58 & 2019-11-03T18:12 &    97.4 & 2.7 \\
12 &     ACE & STEREO-B & 1.01 & 1.07 & 2007-06-08T05:45 & 2007-06-09T05:15 & 2007-06-08T05:45 & 2007-06-09T08:14 &   100.0 & 4.2 \\
13 & STEREO-B & STEREO-A & 1.06 & 0.96 & 2007-05-22T04:26 & 2007-05-23T03:52 & 2007-05-22T14:00 & 2007-05-23T13:30 &    59.2 & 9.1 \\
14 &     ACE & STEREO-A & 1.01 & 0.96 & 2007-05-22T05:08 & 2007-05-23T13:30 & 2007-05-22T14:00 & 2007-05-23T13:30 &    72.6 & 6.0 \\
15 &    STEREO-A &      ACE & 0.97 & 1.00 & 2007-03-29T13:29 & 2007-03-29T20:37 & 2007-03-29T15:04 & 2007-03-30T00:03 &    77.9 & 2.1 \\
16 &     ACE & STEREO-B & 1.00 & 1.02 & 2007-03-29T15:04 & 2007-03-30T00:03 & 2007-03-29T18:24 & 2007-03-30T03:54 &    62.8 & 0.7 \\
17 &      ACE & STEREO-A & 0.98 & 0.97 & 2007-01-15T21:55 & 2007-01-16T03:50 & 2007-01-15T22:16 & 2007-01-16T03:01 &    94.3 & 0.7 \\
18 & STEREO-A & STEREO-B & 0.97 & 0.98 & 2007-01-14T11:57 & 2007-01-15T08:21 & 2007-01-14T12:23 & 2007-01-15T07:55 &    97.8 & 0.1 \\
19 &     ACE & STEREO-B & 0.98 & 0.98 & 2007-01-14T11:46 & 2007-01-15T07:37 & 2007-01-14T12:10 & 2007-01-15T07:37 &    98.0 & 0.3 \\

\bottomrule
\end{tabular}

\caption{Catalog of the 19 events with simultaneous spacecraft measurements. From left to right: Times of entry into the ME of the first and last spacecraft, distances from the Sun for both spacecraft (in au), start and end time for both spacecraft (in YYYY-MM-DDThh:mm format), overlapping time expressed as \% of the ME duration at SC1 and angular separation (in \ang{}).}
\label{tab:boundaries}
\end{table}

All the MEs in Table \ref{tab:boundaries} are probed approximately at the same time by the two spacecraft. 
The results presented in Section \ref{sec:disc_ME} are focused on overall properties (such as the ME mean magnetic field) that are the least affected by the accuracy of the defined boundaries. The table also lists the overlapping duration, \ie\ the duration during which both spacecraft are simultaneously inside the ME, as a percentage of the ME duration at SC1, which ranges from 18\% to 100\% (median of 73\%). The table also gives the distance from the Sun of both spacecraft (in au), and finally the angular separation (in \ang{}), which includes both the longitudinal and latitudinal separations. Depending on the value of the overlap, different studies can be performed as detailed in Section \ref{sec:discussion}.

\begin{table}[h]
\centering
\begin{tabular}{llcccccccc}
\toprule
 \# &                           ID & Bmean1 & Bmean2 & Bmax1 & Bmax2 & DiP1 & DiP2 & ratio1 & ratio2 \\
 & & [nT] & [nT] & [nT] & [nT]& & & &  \\ 
\midrule
   1 & SolarOrbiter\_Wind\_2021-11-04 &   18.2 &   14.5 &  19.2 &  16.1 & 0.50 & 0.49 &    1.0 &    1.0 \\
   2 &      Helios-2\_IMP\_1978-01-04 &   17.2 &   14.4 &  21.0 &  16.4 & 0.47 & 0.54 &    1.5 &    1.0 \\
   3 &      IMP\_Helios-2\_1978-01-06 &   12.1 &   13.2 &  14.9 &  15.3 & 0.48 & 0.47 &    1.3 &    1.6 \\
   4 &      Helios-1\_IMP\_1975-11-17 &   12.8 &   14.3 &  14.5 &  15.4 & 0.58 & 0.49 &    0.7 &    1.1 \\
   5 &         Wind\_NEAR\_1997-12-10 &   11.7 &    8.8 &  16.3 &  13.1 & 0.40 & 0.42 &    2.1 &    2.1 \\
   6 &         Wind\_Juno\_2011-09-17 &   10.3 &   11.5 &  14.1 &  15.0 & 0.40 & 0.48 &    2.8 &    1.4 \\
   7 &         Wind\_Juno\_2011-09-26 &    6.3 &    7.7 &  13.9 &  14.7 & 0.41 & 0.37 &    2.4 &    1.9 \\
   8 &         Wind\_Juno\_2011-10-05 &   10.5 &    8.3 &  14.8 &  11.1 & 0.44 & 0.47 &    1.8 &    1.6 \\
   9 &         Wind\_Juno\_2013-10-30 &    8.9 &    7.6 &  10.9 &   9.3 & 0.48 & 0.47 &    1.0 &    1.3 \\
  10 &         Wind\_Juno\_2013-11-11 &    7.2 &    7.8 &   7.5 &   8.6 & 0.51 & 0.49 &    0.9 &    1.2 \\
  11 &      STEREO-A\_PSP\_2019-11-02 &    6.4 &    6.6 &   7.4 &   8.4 & 0.52 & 0.56 &    0.8 &    0.7 \\
  12 &      ACE\_STEREO-B\_2007-06-08 &    7.6 &    7.3 &   9.7 &  11.7 & 0.50 & 0.54 &    0.9 &    0.7 \\
  13 & STEREO-B\_STEREO-A\_2007-05-22 &   12.5 &    9.6 &  17.6 &  11.8 & 0.44 & 0.56 &    1.0 &    0.8 \\
  14 &      ACE\_STEREO-A\_2007-05-22 &    9.2 &    9.6 &  16.4 &  11.8 & 0.45 & 0.56 &    1.3 &    0.8 \\
  15 &      STEREO-A\_ACE\_2007-03-29 &    4.6 &    4.7 &   5.9 &   6.3 & 0.59 & 0.57 &    0.5 &    0.6 \\
  16 &      ACE\_STEREO-B\_2007-03-29 &    4.7 &    4.3 &   6.3 &   5.8 & 0.57 & 0.58 &    0.6 &    0.5 \\
  17 &      ACE\_STEREO-A\_2007-01-15 &    7.7 &    7.4 &   9.6 &   8.9 & 0.45 & 0.46 &    1.2 &    1.4 \\
  18 & STEREO-A\_STEREO-B\_2007-01-14 &   11.5 &   11.5 &  14.9 &  14.7 & 0.56 & 0.57 &    0.5 &    0.4 \\
  19 &      ACE\_STEREO-B\_2007-01-14 &   11.4 &   11.4 &  14.8 &  14.7 & 0.56 & 0.57 &    0.5 &    0.4 \\
\bottomrule
\end{tabular}

\caption{ME properties measured by the two spacecraft. From left to right, the columns show the event ID (date at which the first spacecraft observed the ME) where the spacecraft are listed in the order they entered the ME, the mean magnetic field inside the ME (in nT), the maximum magnetic field strength measured within the ME, the distortion parameter (DiP) and the front to back ratio. Labels 1 and 2 correspond to the first and second spacecraft to enter the ME, respectively.}
\label{tab:summary}
\end{table}

In Table~\ref{tab:summary}, from left to right, we list the event ID, the mean magnetic field (in nT), the maximum magnetic field (in nT), the distortion parameter \citep[DiP; see][]{nieves-chinchilla2018}, and the front-to-back ratio observed at both spacecraft. The distortion parameter quantifies the asymmetry of the magnetic field profile by giving the location within the ME where half of total magnetic flux is reached. A value of DiP = 0.5 indicates a symmetric ME for this particular measure of asymmetry. A DiP lower than 0.5 corresponds to a ME with a magnetic field stronger in the first half of its profile, and vice versa. Such a feature is often interpreted as the result of a compression of the ME structure \cite[see \textit{e.g.,}][]{masias-meza2016,janvier2019}. %It then gives information about the asymmetry of the magnetic field profile.
We define the front-to-back ratio as the ratio of the ME magnetic field strength averaged over the front 10\% (by duration) divided by the back 10\% (by duration) of the ME magnetic field profile.

%\begin{table}[h]
%\centering
%\include{summary}
%\caption{ME properties measured by the two spacecraft and their positions.  From left to right, the columns show the event ID where the spacecraft are listed in the order they entered the ME, the average magnetic field inside the ME (in nT), the distortion parameter (DiP), the duration (in hours) of the ME, the radial distance (in au) of the spacecraft, the time difference  $\Delta t$ (in hour) between the beginning of the ME at both spacecraft and the angular separation between the spacecraft (in \ang{}). Label 1 corresponds to the first spacecraft to enter the ME.}
%\end{table}

\section{Comparison of the ME properties between the two spacecraft}
\label{sec:disc_ME}

Taking advantage of the proximity of two spacecraft observing the same CME, we now show how different the ME properties can be for angular separation less than \ang{10} and radial separation less than 0.2 au. We first describe an example of a simultaneous measurement event and then generalize the method used on this event to the 19 events present in our catalog.

\subsection{Wind-Juno ME of 2011 September 17}
\label{sec:Wind-Juno}

\begin{figure}[ht]
    \centering
    \includegraphics[width=.9\textwidth]{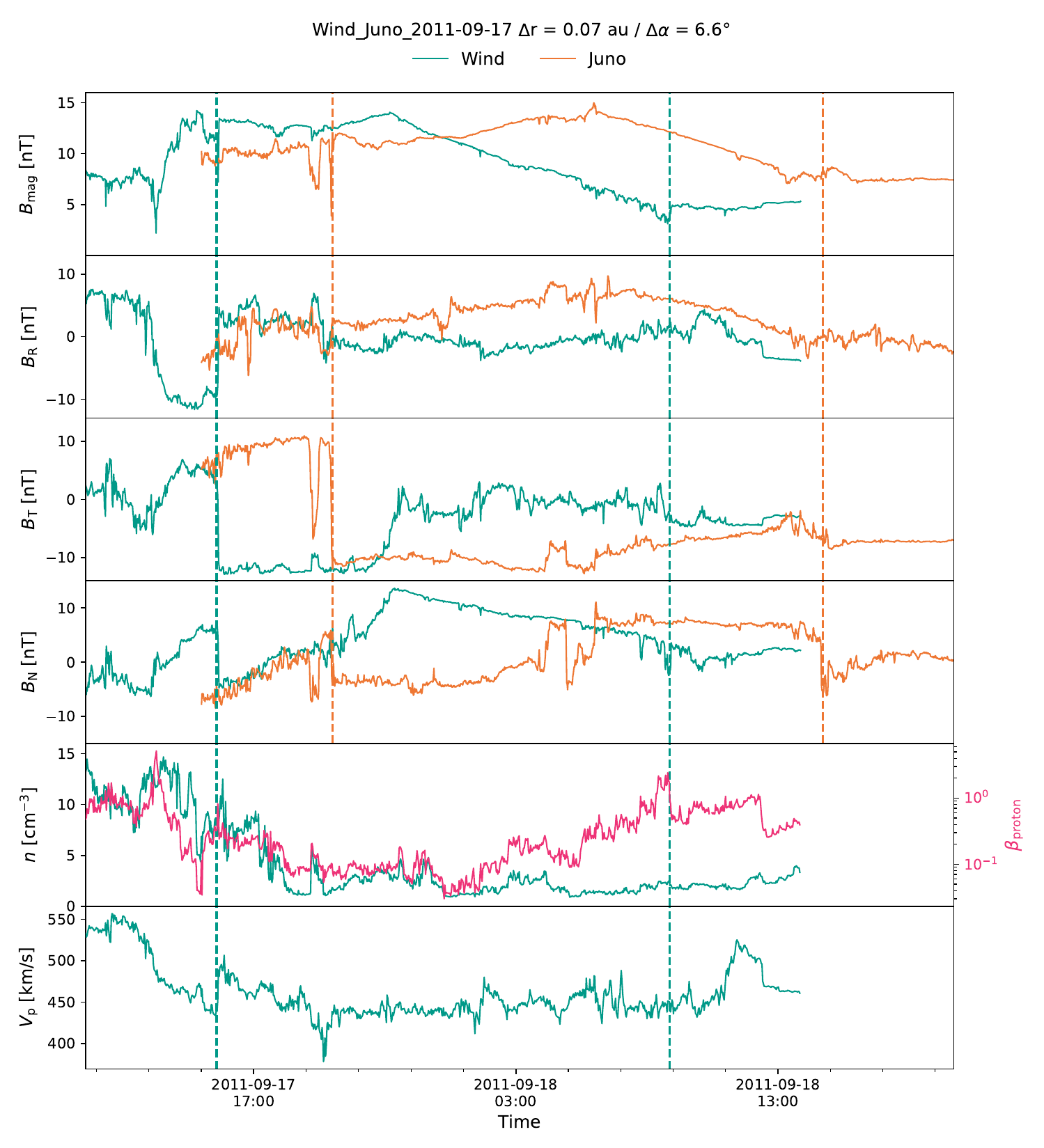}
    \caption{ \textit{In situ} measurements of the 2011 September 17 CME observed at Wind (green) and Juno (orange). The panels show from top to bottom the magnitude of the magnetic field, and its components in RTN coordinates in nT, the density in cm${}^{-3}$, the proton $\beta$ parameter in red and the speed in km.s${}^{-1}$.}
    \label{fig:event_142}
\end{figure}

%It is interesting to note that the same ICME observed almost at the same time shows a difference in asymmetry of 0.03 when the two spacecraft are few degrees apart.

Figure \ref{fig:event_142} presents the magnetic and plasma data for one of the CMEs studied in \cite{davies2021b,davies2022}. The green color corresponds to Wind measurements and the orange to Juno. The ME reached Wind on 2011 September 17. The two spacecraft had a radial separation of 0.07 au, with Juno around 1.07~au, and an angular separation of \ang{6.6}.
At this angular separation, we expect some significant differences, especially in the magnetic field components based on previous studies \citep{kilpua2011,lugaz2018}. This is a relatively slow ME with a speed of $\sim 440$~km\,s$^{-1}$ and no radial expansion, as seen from the velocity measurements observed by Wind presented in the last panel. The ME is preceded by a shock (impacting Wind at 02:57 UT) and a dense sheath region (not shown here). We observe clear shock signatures in the \is\ measurements both at Wind and at Juno. The beginning of the ME is highlighted by the sudden decrease of $B_{\rm T}$ and $B_{\rm N}$ and the increase of $B_{\rm R}$ at both spacecraft. The end of the ME is chosen to correspond to the time when the magnitude of the magnetic field increases suddenly at both spacecraft after an almost monotonic decrease.

The total magnetic field strength profile of the ME has a bell-like shape similar to  what would be expected in typical force-free magnetic cloud models \citep{lundquist1951,marubashi1986,vandas2003}. However, the position of the maximum of the magnetic field within the ME is different at Wind and at Juno: the maximum is closer to the ME leading edge at Wind than it is at Juno. This is also reflected by the DiP which shows a very asymmetric profile at Wind (DiP of 0.40) and a nearly symmetric profile at Juno (DiP of 0.48). Because the decrease of the magnetic field strength in the back of the ME is larger at Wind than at Juno, the average magnetic field increases with heliocentric distance: from 10.3 nT at Wind to 11.5 nT at Juno. However, the peak magnetic field strength is quite similar at both spacecraft.

Different physical mechanisms could be responsible for the shift of the peak towards the center of the ME from Wind to Juno. As already discussed in previous studies \citep[e.g., see][]{farrugia1993,janvier2019}, the magnetic field profile of MEs tends to become more symmetric as the MEs propagate away from the Sun due to relaxation towards a force-free state. However, here, this effect would need to take place over just a few hours (4.4 hours between the beginning of the ME at the two spacecraft), so relaxation cannot explain the changes within such short timescales.
Indeed, in a magnetized plasma, the maximum speed at which information can propagate is at the fast magnetosonic speed ($\sqrt{v_a^2 + c_s^2}$) with $v_a$ and $c_s$ the Alfv\'en speed and the speed of sound respectively. For a ME relaxation to occur, we expect that waves in the ME has enough time to propagage in the entire ME size at least once. The ME size is 0.25~au using the average speed of the ME at Wind while the average fast magnetosonic speed is 153 km\,s$^{-1}$ in the ME at Wind. 
 Within the 4.4 hours delay between the ME front at Wind and Juno, waves have only time to travel about 0.02 au. This corresponds to a very small portion of the ME. Thus, we conclude that relaxation of a large-scale transients such as a CME does not have time to occur from Wind to Juno since information does not have time to propagate far enough.

From the Wind plasma measurements, there is no indication that a fast or dense solar wind stream is following this ME. 
We note that the two spacecraft are \ang{6.6} (0.12 au of arc-length) apart, so this difference in the asymmetry of the magnetic field could be due to inhomogeneous and thus denser solar wind material at Wind than at Juno that locally compresses the magnetic structure. To verify this statement one would need the plasma data at Juno which is not available at that time.
We also want to point out that following the linear force free paradigm (thus with an invariant axis), we cannot explain the shift of the peak of the magnetic field strength profile (even considering different ME inclinations) from Wind to Juno and the differences we observe in the magnetic field components with an angular separation of \ang{6.6} between the spacecraft.

The particular ME presented in this section shows that an angular separation of \ang{6.6} between spacecraft is enough to result in significant differences in the mean magnetic field strength and the asymmetry of its profile. We now extend this comparison to the 19 CMEs presented in our catalog.

\subsection{Magnetic Field Strength}

\begin{figure}[ht]
    \centering
    \includegraphics[width=.8\textwidth]{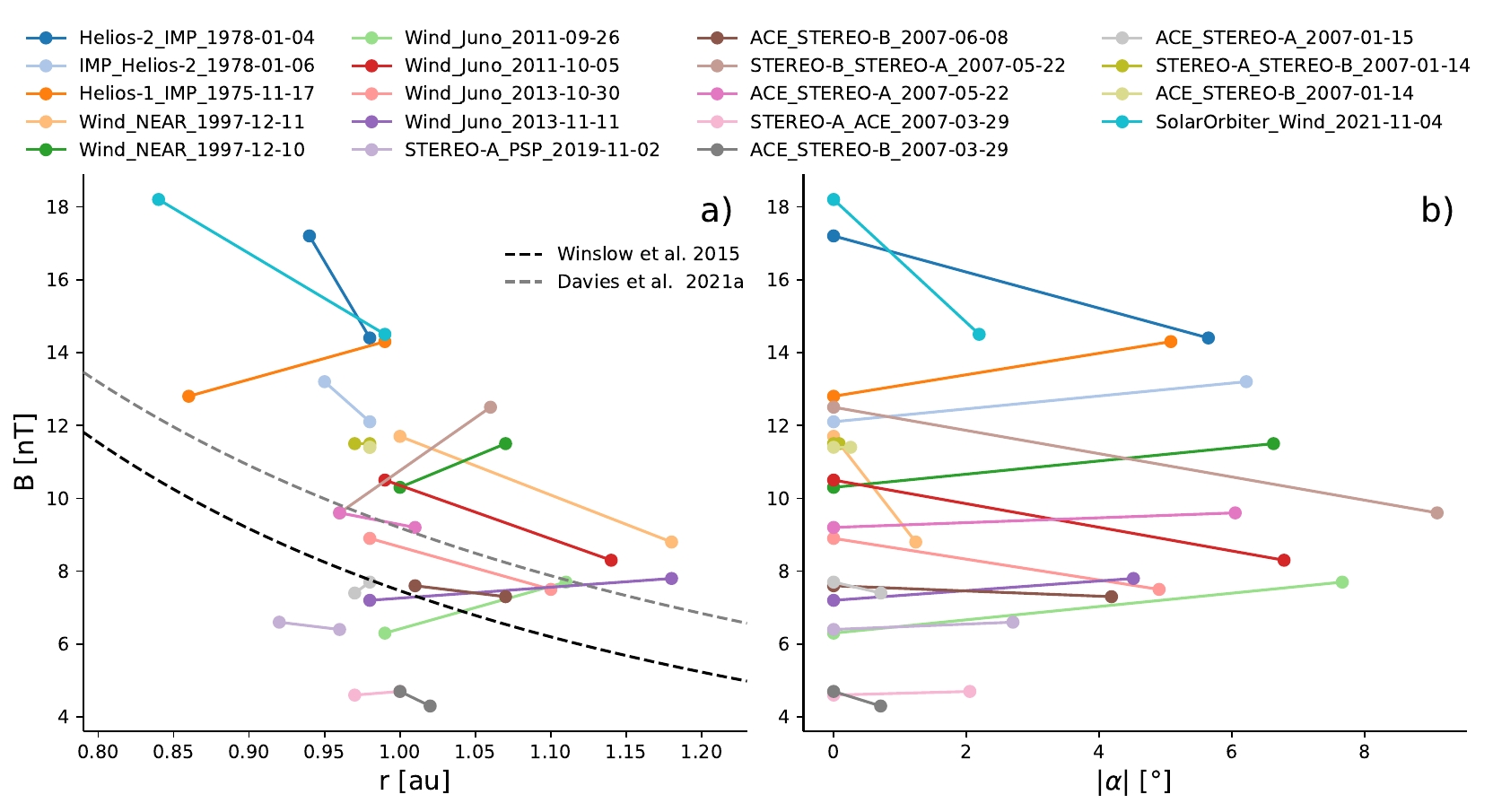}
    \caption{
        The mean magnetic field (in nT) of the ME as a function of the radial distance (in au, panel a) and angular separation (in \ang{}, panel b).
        }
    \label{fig:scatter_plot}
\end{figure}

For the ME events in our catalog, the two spacecraft are separated by up to \ang{10} and 0.2~au and thus the spacecraft may sample different parts of the ME at the same time, and observe significant differences in the mean magnetic field strength. Figure \ref{fig:scatter_plot} shows scatter plots of the mean magnetic field strength of the ME as a function of the radial distance and the absolute angular separation ($\alpha$), respectively. The two observation points of the same ME are connected with a line. Each event is plotted using a different color.

As MEs propagate in the heliosphere, their magnetic field strength decreases \citep{bothmer1998,wang2005,liu2005,richardson2014,winslow2015,lugaz2020a,davies2021b}. This decrease is typically fitted by a power-law with radial distance with an index between $-1$ to $-2$. However, if we study the changes of the ME magnetic field strength on shorter length scales (and shorter time scales), this is not what we observe all the time. The black and grey dashed lines in panel a) show past power law relationships of the magnetic field strength as a function of the radial distance deduced in \cite{winslow2015} and \cite{davies2021b}, respectively, with $B$ the magnetic field strength and $r$ the heliocentric distance. We take the two following evolution laws of the mean magnetic field as a function of the heliocentric distance; $B = 7.46r^{-1.95}$ from \cite{winslow2015} and $B = 9.19r^{-1.62}$ from \cite{davies2021b}. Changes of the ME properties that differ from these power laws have also been discussed for single events observed by multiple spacecraft at different distances from the Sun \citep{good2019,salman2020a,davies2022}. Here, for 6 cases, the average magnetic field is actually found to be larger at the spacecraft taking the measurements at a further radial distance. This highlights that, for radial separations of 0.04--0.2 au between two spacecraft, the changes in the measured ME properties can be significant and may not be primarily due to the ME propagation. This can be explained as either due to i) the effects of the different crossing distances from the ME axis within the paradigm of MEs as cylindrical flux ropes with a straight axis, ii) the effects of different crossings through a more complex and distorted ME, for example without a straight axis, or iii) due to physical processes happening within shorter timescales than the Sun-to-Earth propagation time, for example, interaction with a following fast solar wind stream.  We refer to the first two processes as being due to the 3D nature of the ME structure.
Note that the variation of the mean magnetic field strength over these small radial separations has large event-to-event variability without a clear overall trend, and therefore no final conclusion about any trend can be reached.

Using Table \ref{tab:summary}, we find that the mean and the median of the absolute difference of the average ME magnetic field is 1.3~nT and 1.0~nT, respectively. This result gives an estimate of the typical difference in the average magnetic field measured by two spacecraft separated by small angular and radial separations near 1~au ($\pm$ \ang{10} and 0.2~au).

%In comparison, \cite{regnault2020} conducted a statistical study of \is\ CME properties at 1 au using ACE data and found that the most probable value of the mean magnetic field in the ME is 7.2~nT, while the median is 9.3~nT. Thus, the change in the mean magnetic field strength between the two spacecraft corresponds to about 17\% of the typical mean ME magnetic field (13\% of the median ME magnetic field).

%Following the classification of \cite{bothmer1997} and \cite{mulligan1998} we identified MEs with a high or low inclination, marked with $\blacktriangle$ and $\blacktriangledown$ respectively in Figure \ref{fig:scatter_plot}. If the ME orientation is not clear then it is marked with a $\bullet$. We identify three highly inclined CME, seven lowly inclined and thus nine for which it is not clear. 
%Out of these seven MEs that have an increasing magnetic field in function of $r$, one is lowly inclined, another one in highly inclined and the rest does not have a clear inclination.}

We note that we only looked at events with angular separations of less than \ang{10} and that there is no clear association between the increase in magnetic field and the angular separation of the two spacecraft. It is possible that our results are inconclusive due to low statistics. It also raises the question of the geometry of MEs and their magnetic coherence, in particular, because we also find significant scatter of the mean ME magnetic field as a function of the angular separation. We focus here solely on the average magnetic field strength, not the magnetic field components. Previous work \citep[e.g., see][]{lugaz2018} has shown that the correlation of the total magnetic field strength between two spacecraft decreases much slower with angular separation than the correlation of the components. 

\subsection{ME Distortion}

\begin{figure}[ht]
    \centering
    \includegraphics[width=.8\textwidth]{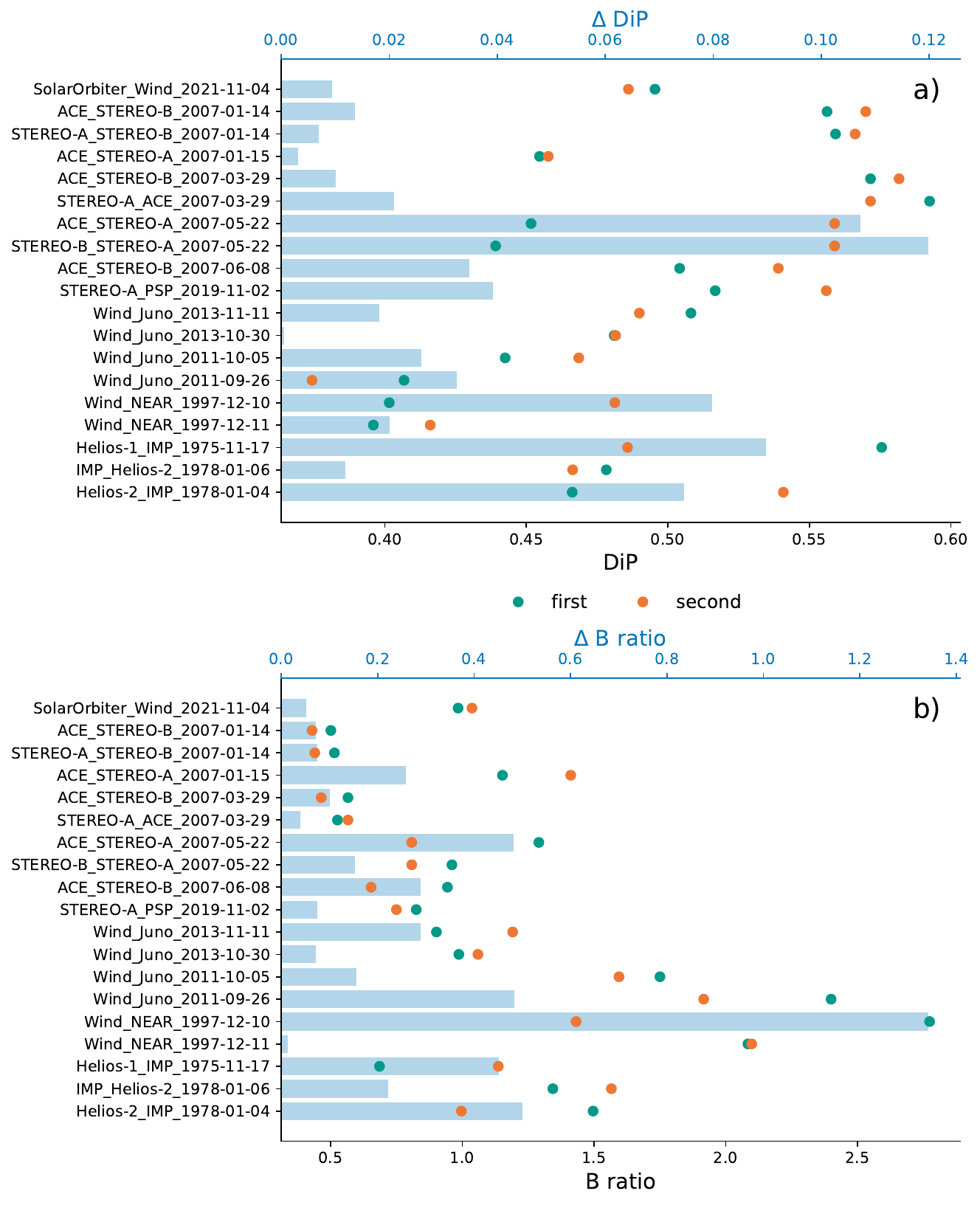}
    \caption{
    Summary plot of the catalog of MEs measured simultaneously by two spacecraft.
    Panels a) and b) show the DiP and front-to-back ratio measured at both spacecraft, respectively, for all the events (green and orange data points, bottom scale) and the absolute value of the difference between the two spacecraft (blue histograms, top scale).   
    }
    \label{fig:summary_plot}
\end{figure}

Figure \ref{fig:summary_plot} summarizes the differences between ME properties measured by each of the two spacecraft for the 19 events under study. Panel a) shows the DiP parameter, and panel b) shows the front-to-back ratio of the magnetic field strength measured at each spacecraft, for each ME measured simultaneously by two spacecraft. 
In the background, the blue histograms show the absolute difference of the DiP (or the front-to-back ratio) between the two spacecraft. 
We use the blue histogram on panel a) of Figure \ref{fig:summary_plot}, to compute the average of the absolute difference of the DiP. This allows us to estimate the typical differences of the asymmetry of the magnetic field for small angular and radial separations near 1~au. The mean value of these differences is 0.04 and the median value is 0.02.

%\soutpars{Next, we turn our attention to the ME duration. Once again, the expectations would be that the ME duration increases with distance away from the Sun. Past studies \citep{bothmer1997,liu2005} have found that the ME radial size increases $\propto r^{0.8}$. Hereafter, we use the duration as a proxy for the size since we do not have plasma measurements at both spacecraft for most of the multi-spacecraft measurements. During such measurements, we expect the ME at both spacecraft to have a very similar duration, which is indeed often the case. However, as shown in Figure~\ref{fig:summary_plot}, the radial and angular separation between the spacecraft do not have a clear effect on the measured ME duration. We find that for four events, the duration decreases by more than 5\% from the closest to the furthest spacecraft from the Sun. Again, this can be explained either through one or a combination of the effect of the impact parameter, the distortion of the ME, or non-radial propagation, the last two of which may be the result of the non-uniform interaction of the ME with the solar wind environment/ transients.}

%
A value of 0.4 for the DiP corresponds to an almost linear magnetic field decrease, with the magnetic field at the front 2.1 times stronger than that at the rear, as shown by Figure \ref{fig:event_140} (Wind\_NEAR\_1997-12-10 case, discussed later in this paper). Such a DiP value corresponds to a strongly asymmetric magnetic profile. The mean difference in DiP of 0.04 between the two spacecraft found for the events in our study, therefore, corresponds to a measurable change in the magnetic field profile asymmetry between two spacecraft taking simultaneous measurements, amounting to about 8--10\% of typical values. This highlights that the differences in the ME properties between the two spacecraft are significant, even for these small separations. 

We find that, for five events, the ME was first measured by the spacecraft that is further from the Sun. This suggests either local distortion of the ME by interaction with the solar wind, significant tilt of the ME front, or non-radial propagation  due to a potential CME deflection. In some cases, it could also be explained if the two spacecraft cross the ME with different impact parameters, but this interpretation would require that the five MEs in this case are inclined since the two spacecraft are separated primarily in longitude, not in latitude.

We compute the Pearson and Spearman correlation coefficients for the DiP and the front-to-back ratio, and we find values of $-0.87$ and $-0.90$, respectively. Therefore, there is a very good anticorrelation between the front-to-back ratio and the DiP parameter in the catalog of simultaneously measured MEs. The front-to-back ratio is thus a good proxy for the overall asymmetry of the magnetic field strength profile. We return to this when investigating the ME instantaneous profile.

%\begin{table}[]
%    \centering
%    \begin{tabular}{c|c|c}
%        Parameter & Mean & Median \\
%         \hline
%         B & 1.1 & 0.95 \\
%         DiP & 0.037 & 0.025 \\
%         Duration & 3.3 & 2.1
%    \end{tabular}
%    \caption{Mean and median of the difference of B, DiP and the duration of the ME. \fr{just a table for now}}
%    \label{tab:mean_med}
%\end{table}
%\subsubsection{Presentation of the cases}

%Figures~\ref{fig:event_140} and \ref{fig:event_142} show the \is\ profiles of both SC for the close conjunction \#140 and \#142. The boundary of the magnetic ejecta is shown in red at Wind and in blue at Juno and NEAR. 

\section{Quantifying the ageing effect and ways to reduce it}
\label{sec:ageing}

In the rest of this study, we take advantage of simultaneous measurements to quantify one of the physical processes that may have a significant effect during the crossing of a CME over a spacecraft: ageing. We define it as the effect of the CME temporal evolution on the \is\ parameter profile during its propagation. Ageing can manifest itself in terms of local processes such as magnetic reconnection, or more global ones, such as a deformation of the cross section occurring during the time when the spacecraft probes the CME. Such phenomena can impact observed CME properties, such as the asymmetry of the magnetic field strength profile. While we do not focus on the ME velocity due to the paucity of measurements, CME deceleration as it passes over a spacecraft is another example of ageing that may affect the quantification of CME expansion.

We look in detail at one ME observed by Wind and NEAR. During the passage of this particular CME, the configuration of the two spacecraft allows us, for the first time, to quantify the ageing effect using the asymmetry of the magnetic field profile in Section \ref{sec:inst_profile}. We then present two methods that aim to reduce this ageing effect and compare the results in Sections \ref{sec:correc} and \ref{sec:reconstruction}
.

\subsection{A ME observed by Wind and Near on 1997 December 10}
\label{sec:Wind-NEAR}

\begin{figure}[ht]
    \centering
    \includegraphics[width=.9\textwidth]{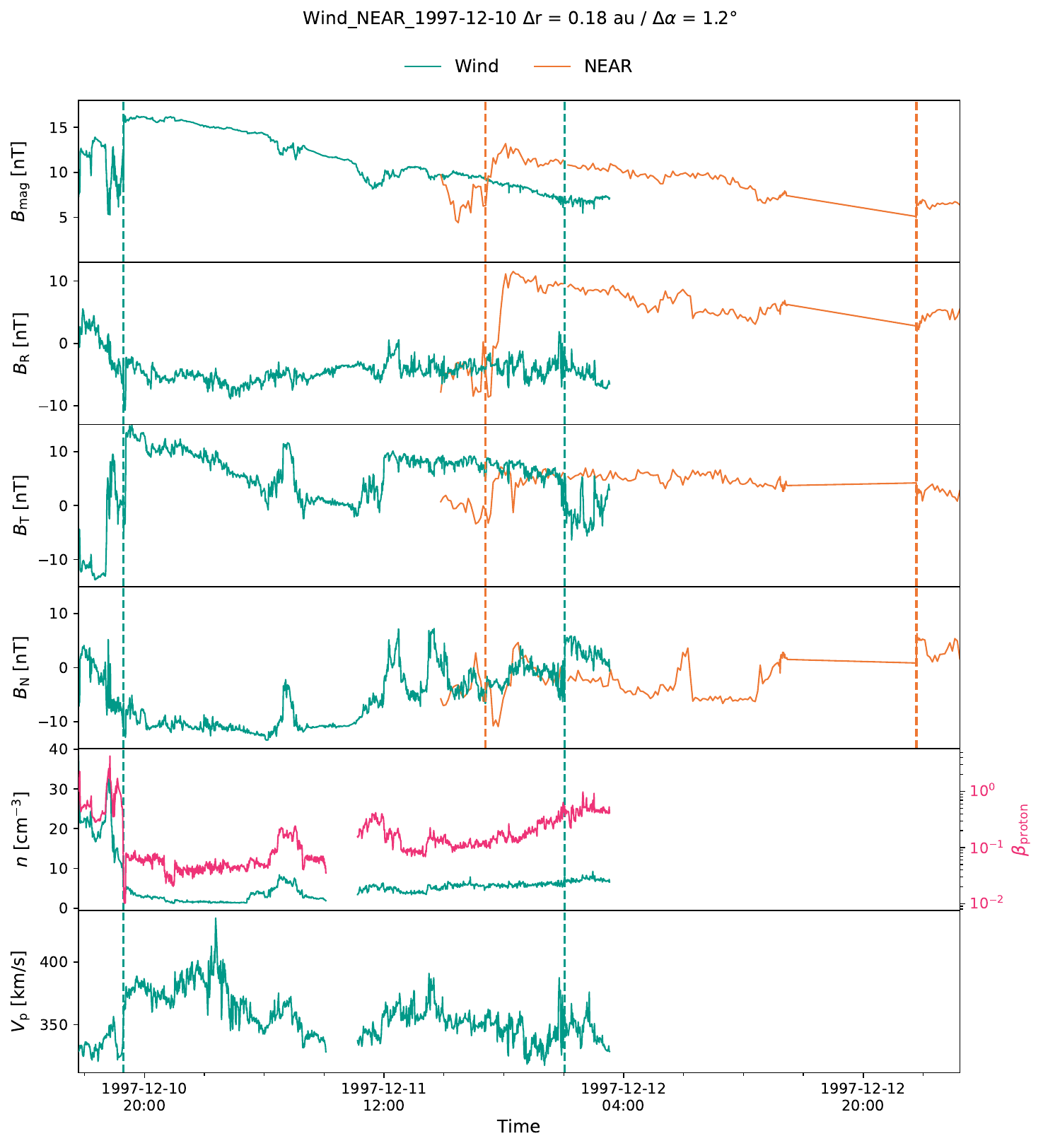}
    \caption{\textit{In situ} measurements of the 1997 December 10 CME observed at Wind (green) and NEAR (orange). Presented in the same format as Figure \ref{fig:event_142}.}
    \label{fig:event_140}
\end{figure}

Figure \ref{fig:event_140} presents the measurements for one of the CMEs studied in \cite{mulligan1999} using NEAR and Wind measurements while these two spacecraft are almost radially aligned ($\Delta \alpha = 1.2$°). The ME reached Wind on 1997 December 10. The radial separation of the two spacecraft was 0.18~au ({\it i.e.} NEAR was at 1.18~au), similar to the typical size of a ME near 1~au.
While both spacecraft measured the magnetic field, plasma measurements are only available at Wind. The ME at Wind is associated with a lower density and lower proton $\beta$ than the nearby solar wind (penultimate panel), clearly highlighting a magnetically-dominated structure. The velocity profile at Wind (last panel) is complex and does not show the clear decreasing profile that is common for most MEs near 1~au. The ME front speed is however slightly higher than the back by about 30~km\,s$^{-1}$, indicating a small radial expansion. The ME is preceded by a shock and a relatively long sheath lasting about 14.5 hours (not shown here). 
Part of the magnetic data was not available when the ME was passing over NEAR. However, there are a few data points at NEAR right before the increase in the total magnetic field strength (due to the increase of $B_{\rm N}$). This feature is also visible at Wind and shows the end of the ME. We thus decide to put the ME end for NEAR at this time. Moreover, we perform a linear interpolation of the missing data at the back of the NEAR ME for the magnetic field strength as it decreases in almost a linear fashion toward the back of the Wind ME.
%We want to point out that this boundary have a impact only on the ME properties presented in Table \ref{tab:summary}.}

More details about the event can be found in \citet{mulligan1999}. Here, our focus is on the simultaneous measurements and what they reveal about the ME temporal evolution during the passage over the spacecraft. 
To do so, we turn our attention to the magnetic field measurements inside the ME. Both spacecraft measure a ME with similar magnetic field components, specifically for $B_T$ and $B_N$. More analysis on the orientation of the ME can be found in \citet{mulligan1999}. We now focus on the magnetic field strength, whose value does not depend on the transformation of the magnetic field components into the same coordinate system and which has been found to be coherent over wider longitudinal separations than the magnetic field components. The temporal profile of the magnetic field strength at Wind shows a strong asymmetry, with almost a linear decrease within the ME. A similar behavior is visible at NEAR. The DiP at Wind is 0.40, whereas it is 0.42 at NEAR.  Such an asymmetry is typically associated with strongly expanding MEs; however, this is not the case for this event, as the ME speed at Wind shows little indication of any radial expansion. This clearly indicates  that radial expansion cannot be the main cause for this asymmetry. The average magnitude of the magnetic field inside the ME is 25\% lower at NEAR as compared with Wind (8.8~nT compared to 11.7~nT). This decrease occurs over a radial distance of 0.18~au, corresponding to a propagation time of 24~hours. This is comparable to the duration of the ME at Wind (29.5 hours). It gives a strong indication that the assumption of the CME as static structure as it passes over the spacecraft is not justified, even in the absence of significant radial expansion, as is the case here. This result suggests that, while propagating from Wind to NEAR, the CME has not expanded significantly but it has aged. This is also consistent with the nearly equal durations at both spacecraft. We return to this point in Section~\ref{sec:inst_profile}.

%Indeed, according to table \ref{tab:summary}, we find that DiP2 (\ie\ at NEAR) value is closer to 0.5 than DiP1 (\ie\ at Wind) with a slightly less magnetized ME at NEAR. This describes a more symmetric ME at NEAR than at Wind.

%\cf{It appears that the B-profile at NEAR is flatter than at WIND....this is one of the expectations from ageing..see above on paper Dynamics of aging....1993}
%\fr{well after such a short time I don't think we would expect much flattening}
%Given the configuration and the distance ($\Delta r$ = 0.18) between the two spacecraft, when Wind is measuring the back of the ME, NEAR is measuring the front of the ME for a window of few hours (\fr{put real value}). See Section \ref{sec:small_overlap} for a potential use of such configuration.

%Assuming that the ICME does not significantly change its magnetic structure between the 2 spacecraft, this number is an estimation of the uncertainty associated with the measurement of the asymmetry of the magnetic structure within \ang{10}. \fr{need help here there is something interesting to say I think with the statistic of the DiP in \cite{nieves-chinchilla2018}. 0.04 is not a small uncertainty basically on the asymmetry of the magnetic field but I am not sure how to say it.}
 
\subsection{ME Instantaneous Asymmetry}
%\subsection{Instantaneous Front-to-Back Ratio: Simultaneous vs. Expansion-corrected}
\label{sec:inst_profile}

In the analysis presented so far, we have not taken full advantage of the fact that the two spacecraft are measuring the ME {\em at the same time} but in different locations. Here, we turn our attention to these simultaneous measurements and how they can help us better understand the true morphology of the MEs.
 
\begin{figure}[ht]
    \centering
    \includegraphics[width=.45\textwidth]{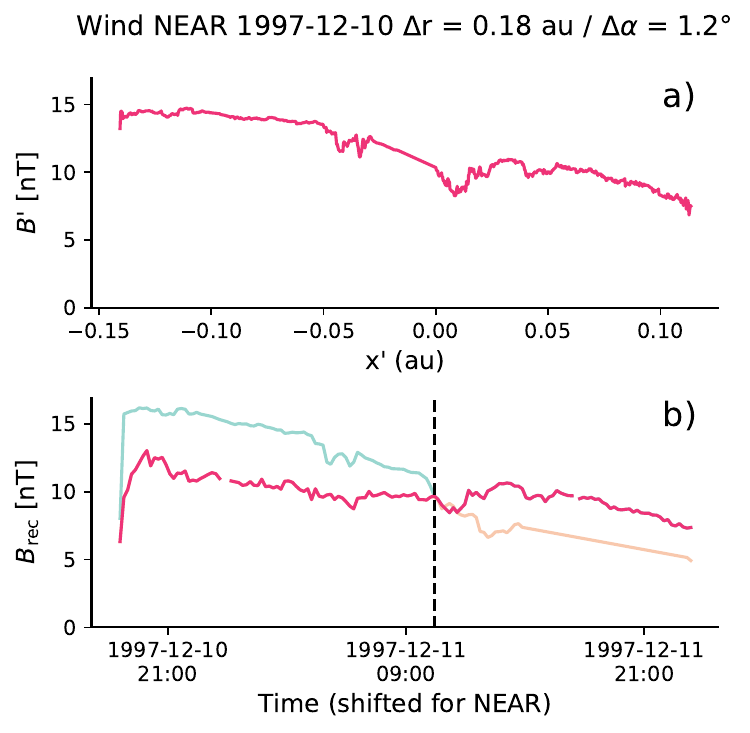}
    \caption{
    Panel a) shows the instantaneous profile following the procedure of \citet{demoulin2020} at Wind on 1997 December 10.
    Panel b) shows the reconstructed profile for the Wind-NEAR event marked by a red line. Light green and orange lines correspond to the portion of the ME profile at Wind and NEAR that are not used to build the reconstructed profile. See Section \ref{sec:reconstruction} for more details.}
    \label{fig:inst_plot}
\end{figure}

In the Wind-NEAR event presented in Figure \ref{fig:event_140}, the spacecraft are in a configuration such that NEAR measures the front of the ME at the exact same time that Wind is measuring the back of the ME, while the angular separation between the spacecraft is very small. We note that as only measurements at the front of the ME at NEAR are used, the definition of the ME end at NEAR does not impact our results. We thus have simultaneous measurements of the magnetic field at the front (probed by NEAR) and the back of the ME  (probed by Wind). With such a configuration, we can then compute an instantaneous front-to-back ratio by combining the CME measurements taken by the two spacecraft, thus  removing any effect of the ME temporal evolution. As discussed in Section~\ref{sec:disc_ME}, the front-to-back ratio is a good proxy for the overall asymmetry of the profile.

Computing the instantaneous front-to-back ratio during the overlapping time gives us $1.4$. The front-to-back ratio of the original profiles is $2.1$ at both Wind and NEAR. We see here that single spacecraft measurements lead us to strongly overestimate the asymmetry of the magnetic field profile.

%$1.42 \pm 0.2$.
%\nl{You need to describe the instantaneous profile, first. How is it built, etc.}

%\nl{I would start two subsections for the two case studies: in each, briefly present the event, spacecraft separations, event speed, previous studies with it, how did you choose the boundaries, etc, this should replace the current structure. Then, discuss the instantaneous profiles}.

\subsection{Correcting for expansion}
\label{sec:correc}
Panel a) of Figure~\ref{fig:inst_plot} depicts the ``instantaneous'' expansion-corrected profile of the ME obtained using the \cite{demoulin2020} methodology for the Wind-NEAR event. This method quantifies the effect of radial expansion on the ME magnetic field. To do so, we perform a linear fit to the speed inside the ME and use the result to compute the parameter $f$ that quantifies the expansion, assuming a self-similar expansion and a constant bulk speed of the CME during the spacecraft crossing. Then, we express the magnetic field profile as a function of spatial coordinates ($B(x)$) and apply the expansion correction. This allows us to plot $B'(x')$ with $B'$ the magnetic field and $x'$ the spatial coordinates, both corrected for radial expansion. We follow this procedure at Wind where solar wind plasma measurements were available. We find that the expansion-corrected front-to-back ratio following the procedure by \citet{demoulin2020} is $1.7$. This shows that most of the asymmetry measured at Wind (where the front-to-back ratio is $2.1$) cannot be explained using the correction of \citet{demoulin2020} alone that assumes self-similar expansion and no changes in the ME properties except for expansion. This result is consistent with the fact that this ME has a small local expansion with, however, large differences between Wind and NEAR in the magnetic field profiles.

%This shows that there is some residual asymmetry after the expansion correction, as observed for the majority of the events presented in \cite{demoulin2020}.

%\cf{Great...Some residual asymmetry.  As was said in paper "Dynamcis of aging..1993, the causes of asymmetry are TWO;
%Aginga nd expansion.  Correcting for AGING leaves us at the mercy of EXPANSION>..Here we are also showing directly that
%indeed expansion des not account for all th asymmetry.}

Next, we compare the two ``instantaneous'' front-to-back ratios of the NEAR-Wind event: the one obtained by taking simultaneous measurements and the one obtained with the instantaneous profile following the procedure of \citet{demoulin2020}. We find a lower asymmetry with the simultaneous front-to-back ratio. This suggests, then, that the \cite{demoulin2020} method does not capture all the asymmetry by correcting for self-similar expansion deduced from the speed. This means that there must be significant ageing of the ME that is independent of the expansion, or that the self-similar assumption does not hold. 
This agrees with the conclusion of \cite{osherovich1993} who pointed out that both expansion as well as ageing must contribute to the asymmetry of the magnetic field profile inside MEs. 
%When comparing the front-to-back ratio of the instantaneous profile with the instantaneous front-to-back ratio, we find that there is some residual asymmetry after the ageing correction, as observed for the majority of the event presented in \cite{demoulin2020}.

\subsection{Reconstruction of the ``True'' Magnetic Profile Inside MEs}
\label{sec:reconstruction}

We now use the simultaneous magnetic profiles of the same ME measured by two spacecraft to build a profile for which the temporal effects are reduced. We assume that, given the small angular separation, each spacecraft probes the same CME portion. Other than that, the reconstructed profile is built relying purely on the data and without making any other assumptions, such as that of self-similarity or lack of ageing beyond the changes due to radial expansion. 

The reconstructed profile is built using the beginning of the measurements at the second spacecraft (SC2) and the end of the measurements at the first spacecraft (SC1). Such a profile should have its time-dependency reduced compared to the profile obtained from a single spacecraft. We switch from the measurements of SC2 to those by SC1 as follows: we find the time closest to the center of the ME as measured by SC2 when the magnetic field strength measurements at SC1 and SC2 are equal. This ensures a continuous reconstructed profile. It is not guaranteed that SC1 and SC2 measure the same magnetic field strength at some point, but it is the case for the event highlighted here.

The result for the Wind-NEAR event is plotted in panel b) of Figure~\ref{fig:inst_plot}. The red profile is the reconstructed profile, while the light green and orange profiles correspond to the original profiles as measured by SC1 and SC2, respectively. We point out that the reconstructed profile only takes the front part of the NEAR magnetic field ME profile and is thus not impacted by the linear interpolation of the back of the NEAR ME.

The front-to-back ratio for the reconstructed profile is $1.5$ for the NEAR-Wind event.
As a reminder, the front-to-back ratio from the spacecraft measurements for the Wind-NEAR event is $2.1$ for both profiles. We thus find here that the reconstructed profile is more symmetric than the measured magnetic field profiles. A similar result is obtained with the DiP values. We find that the DiP is 0.48 for the reconstructed Wind-NEAR profile. This result is expected with the strong correlation previously found between the front-to-back ratio and DiP. 
The profile obtained from the Wind-NEAR event is then significantly flatter than that measured directly or as compared to the ``instantaneous'' profile following the procedure of \citet{demoulin2020} for which the front-to-back ratio is 1.7. 

As this reconstruction only reduces the time-dependency of the measurements at most by half, it is possible that the true magnetic field profile of the ME (meaning a ``cut'' through the ME at a given time) seen by Wind and NEAR would be flat, which is extremely different from the CME profiles observed by each of the two spacecraft.

In spite of its simplicity, this method allows us to reduce the ageing effect of the ME probed by Wind and NEAR since the asymmetry is lower in both reconstructed profiles. However, this approach could be improved by adding geometrical constraints in order to find a more appropriate time when to switch from one \is\ profile to the other.
Overall, the reduction of the ME magnetic field asymmetry results from partially correcting for the effect of ageing (the time-dependent evolution of the ME which differs from radial expansion),
which has been reduced by performing such a reconstruction.

This technique can be applied when two spacecraft are radially aligned. The exact limit of the angular separation that the ``simultaneous measurements" technique requires needs more cases to be firmly established. From the Wind-Juno case discussed in Section \ref{sec:Wind-Juno}, it appears that a radial separation of 0.07 au, and an angular separation of 6.6$^\circ$ (corresponding to an arc of 0.12 au) corresponds to a case when spatial/angular variations dominate over ageing.
Moreover, CMEs can propagate up to 10° in an off-radial direction \citep{al-haddad2022}. Thus, any non-radial flows (even small) would change what we would define as a good ``radial alignment".
That said, even with a perfect radial alignment, the position of the spacecraft with respect to the bulk motion of the CME is still unknown and cannot be established using our limited measurements. This problem is further complicated by the presence of expansion and flows, radial and non-radial, inside the CME. A dedicated cluster of spacecraft that are radially and angularly aligned would provide measurements that will help disentangle those effects.

%\cf{Because the contribution to asymmetry of aging has been removed...should be in abstract?}

%\begin{figure}
%     \centering
%     \begin{subfigure}[b]{.48\textwidth}
%         \centering
%         \includegraphics[width=\textwidth]{figures/134_NEAR_W_1997-12-10.pdf}
%         \caption{$y=x$}
%         \label{fig:y equals x}
%     \end{subfigure}
%     \hfill
%     \begin{subfigure}[b]{0.48\textwidth}
%         \centering
%         \includegraphics[width=\textwidth]{figures/136_Juno_W_2011-09-17.pdf}
%         \caption{$y=3sinx$}
%         \label{fig:three sin x}
%     \end{subfigure}
%\end{figure}

%\begin{figure}[ht]
%    \centering
%    \includegraphics[width=.8\textwidth]{figures/close_conjunction.pdf}
%    \caption{Caption}
%    \label{fig:my_label}
%\end{figure}

%Depending on the overlapping of SC profile, we can do different kind of study : 
%\begin{itemize}
%    \item small overlap simultaneous measurement of the front and back of the ICME 
%    \item medium merging the profile / 2 profiles Lundquist fit
%\end{itemize}

%\subsection{small overlap}
%\label{sec:small_overlap}
%
%- Using the true front-to-back ratio of event \#140. We have a good SC configuration to be able to probe the front and the rear of the ICME at the same time.
%
%\subsection{medium / large overlap}
%
%- Reconstructed profile to reduce the time dependency of the profile.

\section{The potential of Simultaneous Measurements} 
\label{sec:discussion}
%\subsection{Summary}

%Our analysis of the NEAR-Wind ME indicates that assuming that radial expansion is the only source of time-dependency of the magnetic field measurement is not adequate. In fact, ME ageing needs to be investigated by going beyond the assumption of self-similar evolution. \cite{demoulin2020} found that a self-similar expansion is the main source of the asymmetry of only 28\% of the magnetic clouds they investigated.

\begin{figure}[ht]
    \centering
    \includegraphics[width=\textwidth]{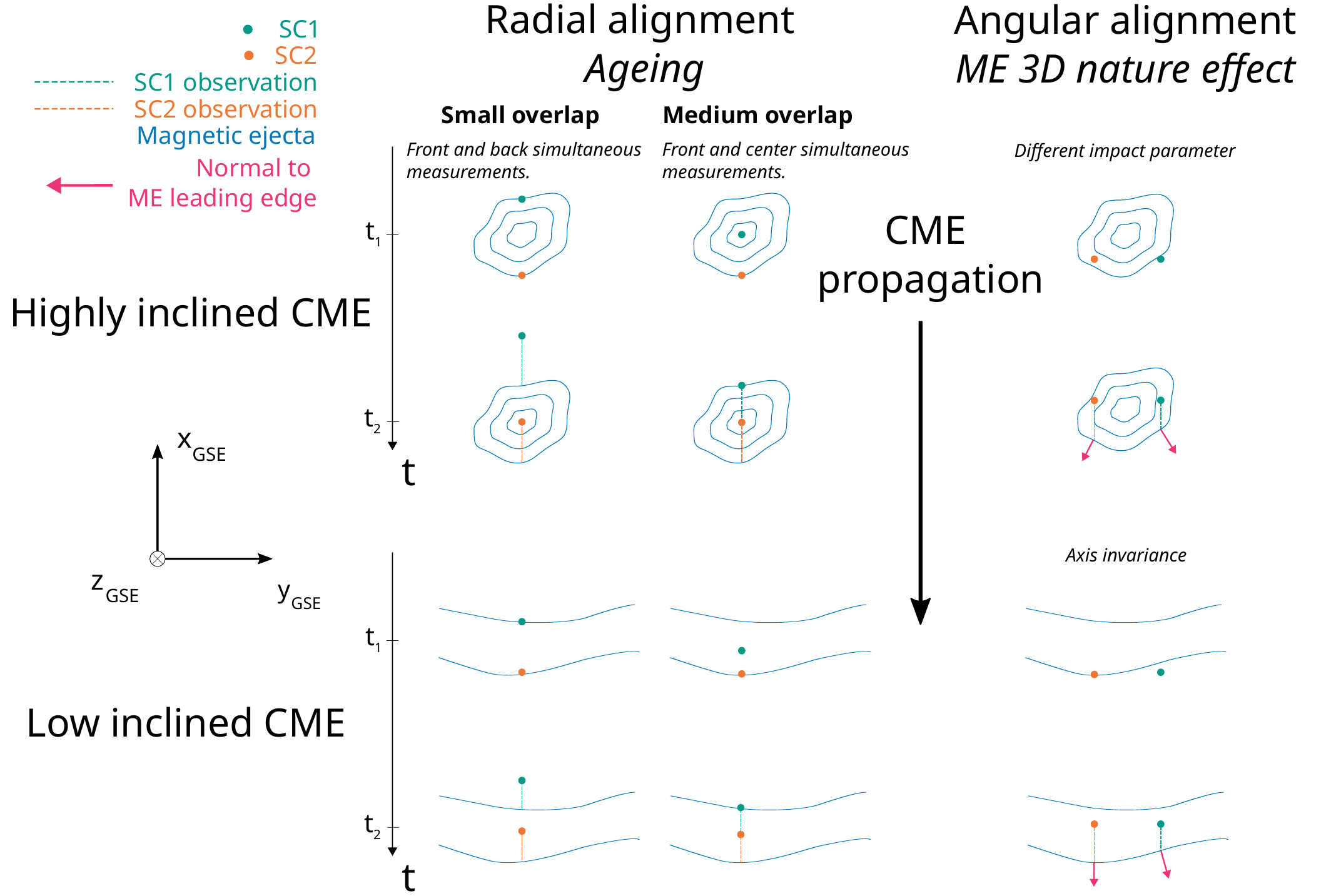}
    \caption{Schematic showing the different possible configurations of the spacecraft during a simultaneous measurement of  a low and highly inclined ME.}
    \label{fig:schematic}
\end{figure}

%\subsection{Potential of Simultaneous Measurements to Better Understand MEs}
%\label{sec:potential}
As shown with our two detailed case studies in Sections \ref{sec:Wind-Juno} and \ref{sec:Wind-NEAR}, simultaneous ME measurements by two spacecraft in close proximity can help us to better understand the ME structure and its spatial and temporal variations. Figure~\ref{fig:schematic} shows a sketch highlighting the different possible configurations for a ME simultaneously measured by two spacecraft, and discusses the different physical processes that can be studied with such configurations. The magnetic structure of the ME represented by the generalized cross-section is in blue. The two spacecraft are in green and orange.

The left column of the figure shows the simultaneous measurement of an ME with radially aligned spacecraft for two different ME orientations. If the two spacecraft are separated approximately by the radial size of the ME (left column), simultaneous measurements of the front and the back of the ME can be obtained to investigate the asymmetry, but there is only limited overlapping time, \ie, during which both spacecraft are inside the ME. This corresponds approximately the situation for the Wind-NEAR event shown in Figure~\ref{fig:event_140}.

If the radial separation between the two spacecraft is smaller than the size of the ME (middle column), both spacecraft are inside the ME at the same time for a longer period of time. The longer the overlap, the more the ageing can be studied and the time-dependency of the profile can be corrected. The configuration for the ``medium overlap'' (separation similar to the ME radial size) as presented in Figure~\ref{fig:schematic} corresponds to the case where the time dependency can be, in theory, divided by two.

The right column of Figure~\ref{fig:schematic} shows the scenario when both spacecraft are approximately at the same heliocentric distance but have some angular separation. In this case, the simultaneous measurements can be used to study the ME 3D nature. In particular, we can test the assumption of axis invariance of the ME, which is a necessary assumption in fitting models \citep[see discussion in][]{al-haddad2019a}. The angular separation that would be optimal for this configuration is still unknown but it is in the range of 2-12$^\circ$ based on our results in this study and previous studies. As highlighted by the Wind-Juno event (see Section \ref{sec:Wind-Juno}), the two spacecraft, in this configuration, should be, as much as possible, at the same radial distance, so that propagation effects between the two spacecraft can be neglected.

\section{Conclusions}
\label{sec:conclusion}

In the present study, we analyze, for the first time, simultaneous measurements of MEs by two spacecraft at different radial distances to investigate the effect of the temporal evolution of the CME as it passes over a spacecraft (ageing), as seen from the measured ME properties. %obtained properties of MEs that were only obtainable through multi-point measurements with spacecraft in close proximity. We 
We start from a database containing measurements of more than two thousand CMEs over four decades to obtain 19  multi-spacecraft simultaneous measurements as the core of the present study. We first compare ME properties probed by two spacecraft when they are close to each other, and then we focus on investigating the effect of ageing on the ME magnetic field strength profile near 1~au.

%\mod{We did not perform any cross-calibration when comparing the magnetic field measurements from the different spacecraft but instrumental effects are expected to be low compared with the differences reported in this study. We also note that the DiP, unlike the magnetic field measurements, should not be affected by any instrumental effects.}

Past statistical studies of CMEs measured at different distances indicate that the ME magnetic field strength decreases relatively steeply ($\propto r^{-\gamma}$ with $\gamma \sim 1.8-1.9$, \citealt{gulisano2010,winslow2015}) with heliocentric distance up to about 1 au. Based on these studies, we would expect the magnetic field strength to decrease by 20--40\% between two spacecraft separated radially by 0.1--0.2~au. This is not what we find here for some events in our catalog. 

One of the key findings of this work is that, in fact, for 6 out of 19 CMEs, the spacecraft further away from the Sun measures the stronger ME mean magnetic field strength. This may indicate that spatial variations, or other phenomena within an ME may result in differences in magnetic field strength of at least 20\%.
Moreover, this catalog allows us to quantify the magnitude of the difference in the ME properties between two spacecraft that measure different parts of the CME at the same time. We find that the ME mean magnetic field varies by 1.3~nT on average, and the DiP parameter by 0.04 between the two spacecraft. This is another key result of the study; it is an estimate of the variability of these parameters when measured by a single spacecraft. It corresponds to a relative difference between 14 and 18\% for the mean magnetic fields when comparing respectively with the average of the most probable and median ME magnetic profile deduced in \cite{regnault2020}.
Given the low number of CMEs in our sample, it is difficult to generalize these findings to all CMEs. However, these shed light on the potential of simultaneous measurements to extend our knowledge about CMEs as discussed in Section \ref{sec:discussion}.

Furthermore, we looked in detail at one specific event for which the two spacecraft (Wind and NEAR) had a very small angular separation and a radial separation similar to the ME radial size. We computed the simultaneous front-to-back ratio of the ME taking advantage of this unique spacecraft configuration. We find that ageing has a significant influence on the magnetic field strength profile of an ME measured by a single spacecraft. We then compared this measure with the ``instantaneous'' magnetic profile deduced using the \cite{demoulin2020} methodology, which is purely based on self-similar expansion using the measured velocity profile. We find that the simultaneous front-to-back ratio is lower than the one computed using the ``instantaneous'' profile. This indicates that self-similar expansion is not the only cause of this ageing.
Results found in this study suggest that one should not to equate ME ageing with expansion as it is clear from this study that these two terms are not equivalent, at least in the case study we present.
%It thus must be investigated in more detail in the future.

Notable spatial variations of the ME properties on an angular extent of \ang{8} have been reported by \cite{hu2022} when using fitting technique with multiple \is\ profile inherent to the model of magnetic structure they used.
Internal magnetic reconnection or interaction with the solar wind can locally change the ME properties and explain these differences. 
In addition, 3D variations within the ME may also contribute to these discrepancies, even for angular separations smaller than \ang{10}, confirming conclusions reached with separations of $\sim 0.5^\circ$ \citep{lugaz2018}. However, such physical processes have not been well studied before, largely due to the lack of dedicated missions (we only found 19 such events over more than four decades of \is\ measurements). Conversely, models also fail to describe accurately the ME structure when the angular separation is too large as shown in \cite{weiss2021a}.

%The catalog \mod{presented in this study} contains 19 separate events measured by two spacecraft within 0.2~au in radial and 10$^\circ$ in angular separation from each other.
%\mod{We find significant changes in the ME average magnetic field strength but also in its asymmetry within these scales.}
%Moreover, we find that ageing processes can affect a ME magnetic field time profile beyond the effects of the ME radial expansion (as pointed out before by \citealt{osherovich1993}). .
Despite our efforts in gathering as many CMEs as possible, we are still limited to 19 events. This low number of events and the lack of readily available plasma measurements taken by spacecraft in our dataset limits the extent and type of investigations that can be done at present. Nevertheless, as shown in this paper, simultaneous multi-spacecraft measurements within CMEs present valuable opportunities to better understand their magnetic structure. Obtaining more simultaneous multi-spacecraft plasma and magnetic field measurements with Solar Orbiter, Parker Solar Probe, STEREO and L1 assets with separations of 0.1--0.2~au, or a dedicated multi-spacecraft mission, should be a priority to understand the temporal and spatial variations within MEs, CMEs and other solar wind structures.

\begin{acknowledgments}

All the data have been are retrieved using a python interface of the Coordinated Data Analysis System (CDAS) web service (\url{https://pypi.org/project/cdasws/}).  The dataset used in this study are AC\_H3\_MFI (1s resolution) and AC\_HO\_SWE (64s resolution) for ACE, HELIOS1\_40SEC\_MAG-PLASMA and HELIOS2\_40SEC\_MAG-PLASMA for Helios-1/2 (40s resolution), SOLO\_L2\_MAG-RTN-NORMAL (0.125s resolution) and SOLO\_L2\_SWA-PAS-GRND-MOM (4s resolution) for Solar Orbiter, PSP\_FLD\_L2\_MAG\_RTN\_1MIN (60s resolution) and PSP\_SWP\_SPC\_L3I (28s resolution) for Parker Solar Probe, STA\_L2\_MAGPLASMA\_1M and STB\_L2\_MAGPLASMA\_1M (60s resolution) for STEREO-A/B, WI\_H0\_MFI (60s resolution) and WI\_H1\_SWE (92s resolution) for Wind, I8\_15SEC\_MAG (15s resolution) for IMP and finally PIONEERVENUS\_MERGED\_SOLAR-WIND\_10M (600s resolution) for PVO, except for the Juno and NEAR spacecraft, for which we use the planetary data system (\url{https://pds.nasa.gov/}).

All authors acknowledge grant 80NSSC20K0431. F.~R., N.~A. and W.~Y. acknowledge grants 80NSSC21K0463  and AGS1954983. N.~L. and F.~R. acknowledge grant 80NSSC20K0700. C.~J.~F. acknowledges support from grants 80NSSC21K0463, and Winds's grant 80NSSC19K1293. E.D. acknowledges NASA grant 80NSSC19K0914 and funding by the European Union (ERC, HELIO4CAST, 101042188). Views and opinions expressed are however those of the author(s) only and do not necessarily reflect those of the European Union or the European Research Council Executive Agency. Neither the European Union nor the granting authority can be held responsible for them.

\end{acknowledgments}

\bibliography{Research.bib}{}
\bibliographystyle{aasjournal}

\end{document}